\documentstyle[12pt,amssymb,epsfig,amsfonts]{article}

\def\e{{\rm e}}

\def\d{\partial}
\def\l{\left(}
\def\r{\right)}

\newcommand{\be}{\begin{equation}}
\newcommand{\ee}{\end{equation}}

\newcommand{\bg}{\begin{gather}}
\newcommand{\eg}{\end{gather}}

\begin{document}
\begin{center}
{\Large \bf
Tunneling into Extra Dimension
and \\ 
\vspace{0.1cm}
High-Energy Violation of Lorentz Invariance}\\
\vspace{0.3cm}
 S.~L.~Dubovsky \\
{\small{\em 
Institute for Nuclear Research of the Russian Academy of Sciences, }}\\
{\small{\em
60th October Anniversary prospect 7a, Moscow 117312, Russia,\\
sergd@ms2.inr.ac.ru
}}
\end{center}
\begin{abstract} 
We consider a class of models with infinite extra dimension, where
bulk space does not possess $SO(1,3)$ invariance, but Lorentz
invariance emerges as an approximate symmetry of the low-energy
effective theory.  In these models, 
the maximum attainable speeds of the graviton,
gauge bosons and scalar particles are automatically equal to each
other and smaller than the maximum speed in the bulk.
Additional fine-tuning is needed in order to assure that the maximum
attainable speed of  fermions takes the same value. A peculiar
feature of our scenario is that there are no truly localized
modes. All four-dimensional particles are resonances with 
finite widths. The latter depends on the {\it energy} of the particle
and is naturally small at low energies.
\end{abstract} 
\section{Introduction and summary}
Physics with extra dimensions has attracted  considerable
interest in the last few years. 
This recent activity is mainly related to the idea that 
large (or even infinite) extra dimensions are possible if matter
fields~\cite{RS,Akama:1982jy,Arkani-Hamed:1998rs} and gravity~\cite{LRS}
are localized on a three-brane embedded in the bulk space of
higher dimension (see, \cite{Volobuev:1986wc} for earlier papers where
the idea of large extra dimensions was discussed).
Hopefully, this approach may provide new insights into the cosmological
constant~\cite{Rubakov:1983bz,Arkani-Hamed:2000eg,Kachru:2000hf} and
hierarchy~\cite{Arkani-Hamed:1998rs,Randall:1999ee} problems.
Another advantage of the scenario with infinite extra dimensions is
that it provides a framework 
 for consistent treatment of those effects which are
difficult to incorporate into conventional four-dimensional
field theories. Examples of such effects are, e.g., modification of
 Newton's law at ultra-large scales~\cite{GRS}, 
electric charge non-conservation~\cite{DRT2} and mass generation for
gauge bosons without the Higgs particle~\cite{ST}.

One more effect, which is hard to incorporate in  conventional
quantum field theories, is violation of Lorentz
invariance at high energies.Certainly, 
Lorentz invariance is broken in our world by the cosmological
expansion of the
Universe. However, this effect
is related to the dynamics at low energies and reveals itself only at
ultra-large distances of the order of 
the Hubble scale. On the other hand,
one may imagine that Lorentz invariance is broken by
high-energy effects in complete theory incorporating quantum gravity.
This idea dates back at least to early 80's
\cite{Nielsen:1983sz,Chadha:1983qq} and since that time both 
laboratory (see, e.g., \cite{Kostelecky:2000fb} for recent review) and
astroparticle (see, e.g., \cite{Bertolami:2000da} for recent discussion)
consequences of high-energy violation of Lorentz invariance we elaborated.

The resulting phenomenology can be addressed by introducing
about 50 Lorentz-violating parameters in the Lagrangian of the
Standard Model~\cite{Colladay:1997iz,Coleman:1999ti}, if one restricts oneself  to
renormalizable CPT even interactions only. The typical experimental
limit on the dimensionless parameters in this approach is at 
the level of $10^{-20}$ while some Lorentz violating parameters are
limited at the level of $10^{-30}$~\cite{Bear:2000cd}.

Consequently, it is of interest to construct a consistent framework
capable of predicting possible relations between various
Lorentz-violating parameters and explaining their smallness. Scenarios
with extra dimensions are natural candidates for such a framework.
Another motivation to consider violation of Lorentz invariance in the
context of models with extra dimensions comes from  cosmology of
these models.  Namely, the standard assumption is that Lorentz
invariance of low-energy effective theory requires  bulk space with
$SO(1,3)$ isometry. This gives rise to an analogue of the flatness
problem in multidimensional cosmology, which may be even more severe
than the flatness problem in the conventional
cosmology~\cite{Chung:2000ji}. The results of this paper indicate that
$SO(1,3)$ isometry of the bulk space may be not necessary.

A mechanism to obtain a low-energy effective theory with approximate
Lorentz invariance from extra dimensions has been recently suggested in
Ref.~\cite{Csaki:2000dm} (see 
Refs.~\cite{Visser:1985qm,earlier,Chung:2000ji} for
earlier works in this direction) where it has been proposed that this
scenario may help to overcome difficulties of 
adjustment mechanisms aimed to solve the cosmological constant
problem.  The basic idea is to consider an asymmetric generalization
of the Randall--Sundrum metric, with different warp factors for
time and space coordinates, 
\be
\label{general}
ds^2=a^2(z)dt^2-b^2(z)d{\bf x}_i^2-dz^2\;.  
\ee 
Explicit constructions
of this metric in Ref.~\cite{Csaki:2000dm} involve (charged) black
hole solutions in the AdS$_5$ space.  It has been assumed that all matter
and gauge fields  reside on the brane located at $z=0$, and, as a
result, the only violation of Lorentz invariance comes from the
gravitational sector.  Approximate four-dimensional gravity emerges at
low energies due to the presence of the localized graviton zero mode
in the background (\ref{general}).  This mode has been found perturbatively,
in the case of small difference between $a(z)$ and $b(z)$. The
complete structure of the low-energy effective gravitational action
remains an open issue.

In this paper we concentrate on a different scenario, namely that
matter and gauge fields, as well as gravitons,
 are localized modes of the bulk quantum fields
in the background
metric of the type (\ref{general}). 
We show that 
violation of Lorentz invariance may reveal itself through 
a rather peculiar
effect at low energies --- metastability of all particles. This effect
takes place if the extra coordinate $z$ is non-compact and if the
ratio of warp factors for time and space coordinates
$a(z)/b(z)$ tends to zero as $z$ tends to infinity\footnote{Note that
both of these properties are absent in the setup of
Ref.~\cite{Csaki:2000dm}.}.  The reason for this effect is that in
this case there are no truly localized modes. Instead,
four-dimensional particles are described by narrow resonances in the
continuum spectrum of the bulk modes.  These quasilocalized modes are
metastable states that decay into the continuum modes. The rate of the
decay depends on the {\it energy} of the particle and is naturally
suppressed at low energies.

From the point of view of a four-dimensional observer,
this type of decays shows up as
literal disappearance of the particles. This
situation is  similar to the metastability of massive particles in
the conventional Randall--Sundrum background studied in
Ref.~\cite{DRT1}.
We will closely follow Ref.~\cite{DRT1} both in spirit and in some of
the technicalities.

This Lorentz-violating effect is related to the presence of the
continuum of  bulk modes in  models with infinite extra
dimensions. It is difficult to incorporate it in the conventional
approach~\cite{Coleman:1999ti}
 based on the notion of the low-energy effective action. 
A natural interpretation may be given in
the holographic language, where disappearance of a particle may be
interpreted as a decay into conformal matter, due to direct (Lorentz
violating in our case) couplings between observed fields and conformal
sector, which corresponds to the bulk modes in the AdS/CFT correspondence. 

The structure of this paper is as follows. In Section~\ref{setup} we
describe a particular class of metrics of the type
(\ref{general}) with $b(z)=const$. To construct these 
metrics as solutions to
the Einstein equations we invoke
an antisymmetric two-form field interacting with gravity.  
These  solutions  may involve
an arbitrary number (including zero) of extra compact warped dimensions. 
These extra dimensions are needed for the localization of
 bosons in the background
metric with $b(z)=const$, their presence is not a necessary feature
of more general Lorentz violating models with quasistable particles.

In Section~\ref{fermions} we describe tunneling into extra dimensions
in the setup of Section 2, by 
using, as a specific example, fermions localized on the brane
by Yukawa-type interactions.
We first find the relation between energy
and spatial momentum of the quasilocalized fermion mode. This
dispersion relation is the same as in  special relativity, but
 with the
speed of massless fermions depending on the value of the Yukawa
coupling between fermions and scalar field responsible for
localization.  We then calculate the life-time of the quasilocalized
fermion and show that it is large at low energies. Both the speed of
the quasilocalized fermion and its life-time do not depend on the
number of the extra compact dimensions. In particular,
the effect of quasilocalization
exists in the absence of extra compact dimensions.

In the end of Section~\ref{fermions}
we present an argument showing
that only {\it quasilocalized} modes can be present in the background 
of
the general type (\ref{general}) with the described above property
($a(z)/b(z)\to 0$ as $z\to \infty$). Thus, with this type of
background metric,
 tunneling into extra
dimensions is a general property of all particles having bulk modes.

In Section~\ref{bosons} we consider
quasilocalized scalar, vector and transverse graviton field, calculate their
dispersion relations and widths, using solutions constructed in Section
\ref{setup} as  concrete examples of Lorentz violating metrics. 
We discuss the fine-tuning conditions needed to
make the maximum attainable speed of fermions being equal to the
speed of photons.

In Section~~\ref{Conc} we briefly review
phenomenological constraints on our scenario following from  cosmic
ray physics and searches for forbidden decays, and present our
conclusions.

\section{Example of the Lorentz violating metrics}
\label{setup}
In our explicit calculations which follow,
we use asymmetrically warped metric of the  form
\begin{equation}
\label{metric}
ds^2=\e^{-2k|z|}\l
dt^2-\sum_{a=1}^{n}(d\theta^a)^2\r-d{\bf x}_i^2-dz^2
\end{equation}
Here $t$ and ${\bf x}_i$ ($i=1,2,3$) are the usual time and space
coordinates and $z$ is a coordinate along an infinite extra
dimension. The coordinates $\theta^a\in [0,2\pi R_a]$ parameterize
internal compact space which we take in the form of
$n$-dimensional torus for the sake of simplicity (more generally,
it can be an arbitrary compact Ricci-flat manifold).  In what follows we 
assume that the radii of compact extra dimensions $R_a$ are smaller
than all distance scales of interest and take all fields to be
constant along the $\theta^a$-directions.

Let us show how 
this metric can be obtained as a solution of field equations.
Consider the following action involving $(5+n)$-dimensional metric $g_{AB}$
and antisymmetric two-form field $B_{AB}$,
\begin{equation}
\label{action}
S=\int d^{5+n}x\!\sqrt{|g|}\l-{M^{n+3}\over
2}R-{\Lambda}+{1\over 4}H_{ABC}H^{ABC}\r\;.
\end{equation}
Here $M$ and $\Lambda$ are bulk Planck mass and cosmological constant
and $H_{ABC}$ is the field 
strength tensor for two-form field $B_{AB}$,
\[
H_{ABC}=\d_AB_{BC}+\d_CB_{AB}+\d_BB_{CA}\;.
\]
In order to obtain the solution (\ref{metric}) we consider bulk space
filled with the
following constant background ``magnetic'' field
\begin{equation}
\label{magnetic}
H_{ijk}=H\epsilon_{ijk}\;.
\end{equation}
The non-vanishing components of the energy-momentum tensor
corresponding to the magnetic field (\ref{magnetic}) in the background
(\ref{metric}) are
\begin{eqnarray}
\label{TEI}
T_{00}=-T_{\theta_a\theta_a}={3\over 2}H^2\e^{-2k|z|},\\
-T_{ii}=T_{zz}=-{3\over 2}H^2\;.
\end{eqnarray}
In addition we assume that there is a source brane located at $z=0$
with the following non-vanishing components of the energy-momentum
tensor
\begin{equation}
\label{source}
\tau_{00}=-\tau_{\theta_a\theta_a}=\sigma_0\delta(z),\
-\tau_{ii}=\sigma_x\delta(z)\;.
\end{equation}
The field equations for the two-form field are satisfied by the Ansatz
(\ref{metric}) and (\ref{magnetic}). The Einstein
 equations lead to
the following set of fine-tuning conditions,
\begin{eqnarray}
\label{Einstein}
2\Lambda=-(n+1)^2k^2M^{n+3},\ H^2=(n+1)k^2M^{n+3},\ \\
\label{sigmas}
\sigma_0=2knM^{n+3},\
\sigma_x=2k(n+1)M^{n+3}\;.
\end{eqnarray}
This fine-tuning is somewhat similar to the fine-tuning inherent in the
original Randall--Sundrum proposal.

A few comments are in order here. First, our solution
may be considered as a limiting case of the hedgehog solutions with
p-form fields described in Ref.~\cite{Gherghetta:2000jf}. The
interpretation is quite different, however --- dimensions which were treated
as small and compact in Ref.~\cite{Gherghetta:2000jf} become
ordinary infinite (or cosmologically large) spatial dimensions in
our setup. On the other hand, small compact extra dimensions $\theta^a$ in the 
metric (\ref{metric}) correspond to the usual spatial coordinates in
the setup of Ref.~\cite{Gherghetta:2000jf}.

Second, the energy-momentum tensor of the
brane involved in our construction violates the positive energy
condition, as is clear  from Eq. (\ref{sigmas}) (the
coefficient $\omega$ in the equation of state $p=\omega\rho$ is
smaller than $-1$ for the ${\bf x}$-component of pressure). Presumably,
this will not lead to an instability as long as ${\mathbb Z}_2$ orbifold
symmetry $z\to -z$ is imposed in analogy to the first Randall--Sundrum
model~\cite{Randall:1999ee}. It is worth  noting that the same
type of exotic sources was involved in the construction of Lorentz-violating
setups in Ref.~\cite{Csaki:2000dm}. It remains to be understood
whether exotic matter violating positive energy condition is a
necessary ingredient of  scenarios with approximate low-energy
Lorentz invariance and localized gravity.

We note finally that, as  pointed out in
Ref.~\cite{Ponton:2001gi}, 
 metric of the type~(\ref{metric}) has an orbifold 
singularity at
$z=\infty$ in the presence of warped extra dimensions $\theta^a$. We
need these extra compact dimensions in order to obtain quasilocalized
bosonic fields (see Section~\ref{bosons}). However, the main effect
studied in our paper (tunneling into extra dimensions in the geometry
of the type~(\ref{general})) is not related to this singularity as
will be demonstrated in the next Section. This effect is present in a
more
general case of non-constant $b(z)$ as well. In that case one need
not add  extra compact dimensions, leading to orbifold singularity, because
graviton and scalar particles are localized by pure gravity, and gauge
field may be localized by some other mechanism (e.g., Dvali--Shifman
mechanism \cite{Dvali:1997xe}). 
Finally, as shown in Ref.~\cite{Ponton:2001gi}, one of possible
 resolutions of this singularity in the context of string
theory preserves the key feature of the Randall--Sundrum model relevant
to the tunneling into extra dimension --- a continuous spectrum of
bulk modes starting from zero.

\section{Tunneling into extra dimension}
\label{fermions}
As the first example of tunneling into extra dimension let us consider
quasilocalized fermions. For concreteness, in  explicit
calculations we make use of 
the metrics constructed in Section~\ref{setup}. In the
end of this Section we show that tunneling into extra dimensions is a
general feature of all particles having bulk modes, provided the Lorentz
violating metrics (\ref{general}) have the properties described in 
Introduction. 

 In the background of the type~(\ref{metric}),
fermionic fields are not localized on the brane by the gravitational
interactions only, for
the same reason as in the original Randall--Sundrum
scenario~\cite{Bajc:2000mh} (see also
\cite{Randjbar-Daemi:2000cr}). Hence, one invokes the localization
mechanism of Refs.~\cite{Jackiw:1976fn,RS}. The simplest setup is as
follows. One considers a domain wall formed by some scalar field
$\phi_k$. This scalar field has a double-well potential with two
degenerate vacua at $\phi_k=\pm v$; the domain wall separates the
region $\phi_k=-v$ at $z<0$ from the region $\phi_k=v$ at $z>0$. In
 flat space, a fermionic field which has a Yukawa coupling to the
scalar, $g\phi_k\bar{\Psi}\Psi$, has an exact zero mode in the domain
wall background. This zero mode is topological and its existence does
not depend on the details of the profile of the scalar field across
the wall. Therefore, it exists also for  infinitely thin wall,
\[
\phi_k(z)=v \mbox{sgn}(z)\;,
\]
which is the case we consider in what follows.

The effective action for fermionic fields which are constant along the compact 
extra dimensions reads (cf. Ref.~\cite{Randjbar-Daemi:2000cr})
\be
S = \int dz\,d^4x\sqrt{|g|}\, \bar\Psi \left( i  
\gamma^{\alpha}\nabla_{\alpha} 
-{ikn\mbox{ sgn}(z)\over 2}\gamma^z+  g\phi_k \right)\Psi, 
\ee
where $\nabla_{\alpha}$ is the spinor covariant derivative with respect to
the metric (\ref{metric}) and $\alpha=0,i,z$. 
For the sake of simplicity we limit ourselves to the case of
massless fermion. The treatment of  massive fermions would require
the introduction of a fermionic doublet in the bulk theory and would
make our formulae more complicated without changing physics
(cf. Ref.~\cite{DRT1}). 
After the change of variables
\[
\Psi=\e^{k(n+1)|z|/2}\tilde{\Psi}
\]
one obtains the following Dirac equation for the 4-spinor
$\tilde{\Psi}$,
\begin{equation}
\left[ E\e^{k|z|} \gamma^{0}+\gamma^{i}p_{i} 
+ \gamma^5 \partial_z - 
g\phi_k(z) \right] \tilde{\Psi} = 0.
\label{dirac}
\end{equation}
which is valid for any number of extra compact dimensions 
(and in
their absence as well).
In order to see that there exists a quasilocalized resonance, let us find
the complex eigenvalue at which there exists a solution to
Eq.~(\ref{dirac}) with the radiation boundary conditions imposed at
$z\to \pm\infty$. It is convenient to choose the following basis for
$\gamma$-matrices,
\begin{eqnarray}
\label{gammas}
\gamma^0=\l
\begin{array}{cc}
0& 1\\
1& 0
\end{array}
\r,\;
\gamma^1=\l
\begin{array}{cc}
i\sigma_2& 0\\
0& -i\sigma_2
\end{array}
\r,\;
\gamma^2=\l
\begin{array}{cc}
0& -1\\
1& 0
\end{array}
\r,\;\nonumber\\
\gamma^3=\l
\begin{array}{cc}
i\sigma_3& 0\\
0& -i\sigma_3
\end{array}
\r,\;
\gamma^5=\l
\begin{array}{cc}
-\sigma_1& 0\\
0& \sigma_1
\end{array}
\r\;\nonumber
\end{eqnarray}
and separate the spinor $\tilde{\Psi}$ into the up and down components
\[
\tilde{\Psi}=\l
\begin{array}{c}
\psi\\
\chi
\end{array}\r\;.
\]
Furthermore, we will make use of invariance of our metric under
spatial 
rotations  and consider a particle moving in
the ${\bf x}^3$-direction, {\it i.e.} choose the spatial momentum in the form
\[
{\bf p}=(0,0,p)
\]
In terms of two-component spinors $\psi$ and $\chi$, 
Eq.~(\ref{dirac}) translates into a
set of coupled equations
\begin{eqnarray}
\label{psi}
E\e^{k|z|}\chi+ip\sigma_3\psi-\sigma_1\d_z\psi-g\phi_k\psi=0\;,\\
\label{chi}
E\e^{k|z|}\psi-ip\sigma_3\chi+\sigma_1\d_z\chi-g\phi_k\chi=0\;.
\end{eqnarray}
After eliminating $\psi$ one obtains the second order equation for
$\chi$,
\be
\label{fShr}
\left[E^2\e^{2k|z|}+\d_z^2-k\mbox{sgn}(z)\l\d_z-
(p\sigma_2+g\phi_k\sigma_1)\r-g\phi_k'\sigma_1-(g\phi_k^2+p^2)\right]\chi=0
\;.
\ee
The differential operator in the left hand side  of Eq.~(\ref{fShr})
coincides with the differential operator entering
the second order equation for massive 
fermions (with mass equal to $p$) in the
Randall--Sundrum background; the latter has been
 considered in
Ref.~\cite{DRT1}. Consequently, we may make use of the result of that
work and find that there is a complex eigenvalue determined by the
following equation at $p\ll gv$,
\[
{H_{M/k-1/2}^{(1)}\l E/k\r\over H_{M/k+1/2}^{(1)}\l E/k\r}={p\over
2gv}\;,
\]
where $M$ is defined in such a way that
\[
gv+ip=M\e^{i\alpha}
\]
with real $M$ and $\alpha$.
We are interested in the case of small energies and momenta $E,p\ll k,gv$.
In this case one expands the Bessel functions at small values of the
argument and finds
\[
E=E_0-{i\Gamma\over 2}
\]
with
\be
\label{fdisp}
E_0=\l 1-{k\over 2gv}\r p
\ee
and
\be
\label{fwidth}
\Gamma={2\pi E_0\over [\Gamma\l
gv/k+1/2\r]^2} \l {E_0\over 2k}\r^{2gv/k-1}\;.
\ee
Equations (\ref{fdisp}) and (\ref{fwidth}) describe a narrow resonance
at low energies, provided
\[
{gv\over k}>{1\over 2}
\]
This resonance can be interpreted as a four dimensional metastable
particle. The $\chi$-component of its wave
function is determined by
Eq. (\ref{fShr}) and $\psi$-component of its wave-function
 may be found then from Eq. (\ref{chi}).

The speed of the quasilocalized massless fermion is equal to
\[
c_f=1-{k\over 2gv}
\]
as is clear from 
Eq. (\ref{fdisp}). This speed depends on the parameter $k/2gv$
and is 
always smaller then the speed of a massless point-like particle tightly
bound to the brane and moving along
$z=0$ (the latter case corresponds to
the limit $gv \to \infty$). This does not necessarily imply that the
maximum speed is not the same for different {\it quasilocalized}
particles. As we will see below, quasilocalized scalar field, photon and
graviton in the background (\ref{metric}) have equal speeds (smaller
than the speed of a tightly bound massless particle). 
In order that the quasilocalized fermion has the same 
speed, additional fine-tuning of  parameters is needed.

Another non-trivial manifestation of the violation of Lorentz invariance
is non-zero imaginary part of the fermion energy, as given by
Eq. (\ref{fwidth}). This imaginary part determines the probability for
quasilocalized fermion to tunnel into extra dimension.
The tunneling probability is small at low energies
and grows with energy, as  follows from Eq.~(\ref{fwidth}).
This effect is present in the absence of extra compact dimensions (at
$n=0$) and, hence,
is not related to the orbifold singularity at  infinity.

Let us now present an argument indicating that metastability of all
particles having bulk modes is a general property of the metrics of
the type (\ref{general}) with $a(z)/b(z)\to 0$ as $z\to \infty$.
Indeed, a mode 
equation for a bulk field $\Phi$ in the background
of this type may be presented
in the following general form
\be
\label{argument}
\left[\Delta(z)+p^2f(z)-E^2\right]\Phi(z)=0\;, \ee where $\Delta(z)$
is a positive semi-definite differential operator and the function
$f(z)=a^2(z)/b^2(z)$ vanishes at infinity.  The key property of models
with infinite extra dimensions, essential for our argument, is that
the operator $\Delta(z)$ has a continuous spectrum of energies
starting from some $E^2=E^2_{min}$ at $p=0$. This property may be
considered as a definition of infinite extra dimension and means that
the energy of the particle related to its motion along the
$z$-direction is not quantized, at least starting from some large
enough value.  Note that in most of known models with infinite extra
dimensions and localized gravity, the continuum spectrum starts from
zero energy\footnote{See, however, Refs.~\cite{Brandhuber:1999hb,ST}.
We thank P.~Tinyakov for pointing out the possibility of non-zero
$E_{min}$.}  $E_{min}=0$. At non-zero $p^2>0$, the term $p^2f(z)$ in
the left hand side of Eq. (\ref{argument}) vanishes at large $z$ and
thus it is irrelevant at infinity. Since the continuum eigenvalues are
determined by the large $z$ asymptotics only, Eq.~(\ref{argument}) has
the same continuum spectrum for all $p^2$ including $p^2=0$.  This
implies that there are no truly localized modes with $E^2>E^2_{min}$
(there are no truly localized bound states embedded in the continuum)
and all four-dimensional particles become unstable above this
energy. A similar argument was presented in Ref.~\cite{DRT1} for
massive fields in the Randall--Sundrum background.

\section{Quasilocalized bosons}
\label{bosons}
As another example of tunneling into extra dimension, we
consider propagation of the scalar field $\phi$ with mass $\mu$
in the background metric (\ref{metric}). The corresponding field
equation  has the following form
\begin{equation}
\label{scalareq}
\left[-\d_z^2+(n+1)k \mbox{sgn}(z)\d_z+\mu^2+p^2-E^2\e^{2k|z|}\right]\phi=0\;,
\end{equation}
where $p$ is the spatial momentum and $E$ is the energy.
This equation is analogous to the field equation  for  massive
scalar field in the Randall--Sundrum background. At a given value of
$p^2>0$, there is a continuum spectrum of modes with arbitrary energies
$E^2>0$ and no localized modes. However, as in the case of massive
fields in the Randall--Sundrum metric, there exists a resonance mode which
describes quasilocalized particle. Indeed, let us show that the mode
equation, Eq.~(\ref{scalareq}), has a complex eigenvalue when the
radiation boundary conditions are imposed at $z\to\pm\infty$.  The
solution to Eq.~(\ref{scalareq}) which satisfies the radiation boundary
conditions is
\begin{equation}
\label{solution}
\phi(z)=c\e^{k(n+1)|z|/2}H^{(1)}_{\nu}\l{E\over k}\e^{k|z|}\r\;,
\end{equation}
where $H^{(1)}_{\nu}(x)$ is the Hankel function, the constant $c$ is
determined by the normalization condition and
\be
\label{nu}
\nu=\sqrt{\l{n+1\over 2}\r^2+\l{p\over k}\r^2+\l{\mu\over k}\r^2}
\ee
The first derivative of $\phi(z)$ should be continuous at $z=0$, as is
clear from Eq.~(\ref{scalareq}), 
\[
\d_z\phi(+0)-\d_z\phi(-0)=0\;.
\]
The latter condition implies the following dispersion relation
\be
\label{dispersion}
{E\over k}{H^{(1)}_{\nu-1}\l{E\over k}\r\over H^{(1)}_{\nu}\l{E\over
k}\r}+{n+1\over 2}-\nu=0\;.
\ee
Let us consider small energies and momenta $\mu,E,p\ll k$. Then expanding
the Bessel function at small argument one finds
\be
\label{complexE}
E=E_0-{i\over 2}\Gamma
\ee
with
\be
\label{realE}
E_0^2={n-1\over n+1}(p^2+\mu^2)
\ee
and
\be
\label{imagE}
\Gamma={2\pi E_0\over\Gamma(n/2)\Gamma(n/2-1)}\l{E_0\over 2k}\r^{n-1}\;.  
\ee 
We see that at $n>1$ (this is the case we consider in the rest of the Section)
the width of the resonance is much smaller than $E_0$. As a
result, this resonance can be interpreted as a four-dimensional
metastable particle. This particle decays through tunneling into
extra dimensions. Violation of Lorentz invariance reveals itself in
the fact that the decay rate grows with the energy of the particle. The
factor of $(n-1)/(n+1)$ in the dispersion relation (\ref{realE}) implies
that the maximum speed of this particle is smaller than the speed of a
tightly bound massless particle moving along  the brane at $z=0$.  
In order that this speed be equal to the
speed of the quasilocalized fermions, the following fine-tuning
condition should be satisfied
\be
\label{cfine}
1-{k\over 2gv}=\sqrt{{n-1\over n+1}}\;.
\ee 
It would be  desirable to obtain this fine-tuning condition as a
consequence of some symmetry in the underlying 
theory. Another solution of this problem may be to construct a
setup where fermions are localized by gravity only, like all other
fields. See, e.g., Refs.~\cite{Randjbar-Daemi:2000cr,Neronov:2001br}
for recent attempts in this direction. Alternatively, one can localize
fermions on the additional spectator brane like in the Lykken--Randall
scenario~\cite{Lykken:2000nb}. This brane should be located in the
place  where the maximum speed of tightly bound
particles in ${\bf x}_i$ directions is the same as the speed of the
quasilocalized bosonic modes.
Finally, one can consider 
metrics of the type~(\ref{general}), i.e., 
 more general than those described in
Section~\ref{setup}. In that case the (approximate) equality between maximum
attainable speeds of fermions and bosons may be attributed to the
small difference between the warp factors $a(z)$ and $b(z)$.

Let us now consider the propagation of a massless vector field $V_A$ in
the background (\ref{metric}).
The field equation   in  curved space has
the form
\begin{equation}
\label{vectoreq}
\d_A\l\sqrt{|g|}F^{AB}\r=0\;,
\end{equation}
where $F_{AB}=\d_A V_B-\d_B V_A$. To classify
propagating modes of the vector field in the background (\ref{metric}), we
find it convenient to work in the gauge 
\[
V_0=0\;.
\] 
In this gauge the $(0)$-component of Eq. (\ref{vectoreq})
takes the following form,
\be
\label{V0}
\d_0\left[ \d_z\l\e^{-(n-1)k|z|}V_z\r+\e^{-(n-1)k|z|} \d_iV_i\right]=0
\ee 
The fact that the combination in the square bracket in the
left hand side of Eq. (\ref{V0}) is constant in time makes it
possible to use the residual gauge freedom  to set this
combination to zero, 
\be
\label{gauge}
\d_z\l\e^{-(n-1)k|z|}V_z\r+\e^{-(n-1)k|z|}
\d_iV_i=0\;.
\ee
Then one can rewrite the $(z)$-component of Eq.
(\ref{vectoreq}) in the following form,
\be
\label{Vz}
\left[-\d_z^2+(n-1)k 
\mbox{sgn}(z)\d_z+p^2-E^2\e^{2k|z|}+2(n-1)k\delta(z)\right]V_z=0\;.
\ee At zero energy and momentum, the field $V_z=const$ ceases to be a
solution of this equation because of the presence of the extra
delta-functional term in the left hand side of Eq.  (\ref{Vz}), which
is absent in Eq.~(\ref{scalareq}).  As a result, Eq. (\ref{Vz})
does not admit localized modes at $E=p=0$ and one has no reason to
expect the presence of narrow resonances at small but
non-vanishing energy
and momenta. Straightforward calculation analogous to 
one outlined in the beginning of this Section
shows that such resonances are indeed
absent.  Consequently, the $z$-component of the vector field (or,
equivalently, the longitudinal component of the photon, see Eq.
(\ref{gauge})) does not have (quasi)localized modes in our setup.

Now, it is straightforward to check that transverse components $V_i$
satisfy the same mode equation as the massless scalar field,
Eq.~(\ref{scalareq})  with $\mu=0$. Consequently, a quasilocalized
transverse photon with the same speed of light as the maximum speed of
the scalar particle exists at $n>1$.

Components $V_{\theta^a}$ of the vector field do not couple to matter as
long as there are no currents along the extra compact
dimensions. However, for the sake of completeness, let us consider the
propagation of these components as well.
Taking ($\theta^a$)-component of Eq. (\ref{vectoreq}) one obtains
\be
\label{Va}
\left[-\d_z^2+(n-1)k
\mbox{sgn}(z)\d_z+p^2-E^2\e^{2k|z|}\right]V_{\theta^a}=0\;.
\ee
Comparing this equation with the equation (\ref{scalareq}) for the
scalar field we see that these components correspond to quasilocalized
scalar particle with the maximum speed equal to $\sqrt{(n-1)/(n-3)}$
at $n>3$. 

Let us now briefly discuss the quasilocalized graviton mode in our
setup. The mode equation for
transverse traceless gravitational waves may be obtained
by considering metric perturbation of the form
\[
\delta ds^2=h(z)h_{ij}({\bf x},t)d{\bf x}^id{\bf x}^j
\]
with
\[
\d_ih_{ij}=\delta^{ij}h_{ij}=0
\]
It is straightforward to check that these perturbations decouple from
 perturbations of other components of the metric and of the
two-form field $B_{AB}$ and satisfy
 the same mode equation as the scalars, Eq.~(\ref{scalareq}). This is the same
situation as in the   Randall--Sundrum 
model~\cite{Giddings:2000mu,Bajc:2000mh} and as in the setup of
Ref.~\cite{Csaki:2000dm}.
Consequently, there exists a
quasilocalized graviton in our setup with 
the same low-energy properties as the massless scalar boson.
We will not study the excitations of other components of the
metric and of the two-form $B_{AB}$ here; this study would be 
necessary to understand  the complete
structure of the low-energy effective action for gravity in our setup
and in the setup of Ref.~\cite{Csaki:2000dm}. We hope to return to
this  issue in future.

We would like to make only a couple of comments in this regard. The
first one is that in
Ref.~\cite{Csaki:2000dm}, there were suggested two possible explanations of
the difference between the speed of the localized graviton and the speed of
a tightly bound massless particle moving along 
the brane at $z=0$. The first possibility is that the
effective action for gravity explicitly breaks the 4D Lorentz
invariance at $z=0$, as different coefficients
for time and space coordinates in the kinetic term of the
graviton are introduced. 
The second possibility is that the effective action 
violates the weak equivalence principle by the presence of some extra
fields which couple differently to matter and gravity. 
This may force gravitational waves to propagate differently
than the tightly localized particle. 

To infer which of these two possibilities is actually realized,
we note
 that the correct value of the maximum speed of
the quasilocalized bosons can be obtained in the following
empiric way. The scalar field action in the background (\ref{metric}) has
the following form for the modes which are constant over compact
coordinates $\theta_a$ 
\be
\label{scalactio}
S_{sc}\propto\int\! dtd{\bf x}dz\l
\e^{-(n-1)k|z|}\dot\phi^2-\e^{-(n+1)k|z|}\l\d_i\phi\r^2-\e^{-(n+1)k|z|}\mu^2\phi^2\r\;.
\ee 
Note that the field equation (\ref{scalareq}) admits
a normalizable mode $\phi(z)=const$ with zero energy when $p=\mu=0$.  
To obtain the effective four-dimensional action for
quasilocalized modes of low energy and momentum, one can plug a
field $\phi(t,{\bf x})$ independent of $z$ in the action
(\ref{scalactio}) and integrate over $z$. Then one will obtain 
precisely
the dispersion relation (\ref{realE}) due to  different
coefficients in front of time and space derivatives in
Eq.~(\ref{scalactio}).

The fact that the above empirical way
to obtain the low energy effective action for the scalar (and
other bosons) gives the correct value for its speed
due to the difference in the coefficients in front of time and space
derivatives suggests that it is the effect of the first type that
takes place (at
least in our setup). Certainly, our observation
 does not exclude the possibility that the second
effect may also be present.

The second comment is that one may worry that our setup suffers from 
a van Dam-Veltman-Zakharov (vDVZ) discontinuity as other theories
with quasistable graviton~\cite{Dvali:2000rv}. In this regard it worth
noting, that, as it was argued recently, this discontinuity is likely
to be an artifact of the linear
approximation~\cite{Deffayet:2001uk}. Secondly, and more relevant for
the class of model we consider, this discontinuity is due to the fact
that quasilocalized graviton has extra polarization states in model discussed
in~\cite{Dvali:2000rv}. In our models, graviton is strictly massless
and may decay not because of non-zero mass 
but due to the presence of bulk modes with smaller
propagation velocities. Consequently, one has no a priori reason 
to expect extra polarizations, which would give rise to the vDVZ
discontinuity. Explicit analysis presented above demonstrates that these
polarizations are indeed absent in the photon case. The analysis of
the linearized gravity in asymmetrically warped spaces which is
necessary to address the same issue for a graviton is beyond the scope
of this work.

\section{Discussion}
\label{Conc}
In this concluding section, let us first briefly discuss
 potential experimental
constraints on the scenario suggested in this paper, by making use of
the brane world
solutions of Section~\ref{setup} as a toy model. 
We assume that
the fine-tuning condition (\ref{cfine}) is satisfied and neglect
possible non-universality in the maximum attainable speed of 
different quasilocalized particles which may appear
due to quantum corrections. The latter issue
requires a separate  study. Then the limits on the
parameters of our model are related to the instability of energetic
particles and come from  searches for forbidden decays and from 
physics of high-energy cosmic rays. In all our estimates we  take
the 
 parameter $k$ to be equal to the Planck mass $M_{Pl}\sim 10^{19}$~GeV.
Certainly, one can make all  limits  weaker by taking larger
values of $k$. Also, one may expect that these constraints are weaker
for more general metrics of the type given in Eq.~(\ref{general}),
if the difference between warp factors $a(z)$ and $b(z)$ is not so
drastic as in  Eq.~(\ref{metric}).

The first limit comes from the stability of  electron.  We estimate
the width of  electron against the decay $e^-\to nothing$ by
plugging $E_0=m_{e}$ in the expression (\ref{fwidth}) for the width
of massless fermion and finding the parameter $gv/k$ from the
fine-tuning condition (\ref{cfine}). As a result one finds 
that for two extra compact dimensions 
(the smallest number for which the localization of bosons takes place in
our toy model), 
the life-time
of  electron is of order 100 years, which is certainly
unacceptable. However, for three extra compact dimensions this
life-time is  $9\cdot 10^{25}$~years which is larger than the
current experimental limit $\tau_{e}>4.2\cdot
10^{24}$~years~\cite{Groom:2000in}.

It is not quite obvious that Eq. (\ref{fwidth}) can be
directly applied for estimating the proton life-time. Proton is a
composite object and this fact may lead to an additional suppression of the
decay rate. A naive application of Eq.~(\ref{fwidth}) with
 $E_0=m_{p}$ excludes our setup with three extra compact
directions, but for four extra compact dimensions gives the value
$\tau_{p}=9.2\cdot 10^{34}$~years which is much larger then the
experimental limit on the decay $p\to nothing$,
$\tau_{p}>5\cdot 10^{26}$~years~\cite{Groom:2000in}.

Another set of limits comes from the high-energy cosmic ray data.
Observation of  $20$~TeV photons from the distant blazar Markarian
501~\cite{mark} with redshift $z\sim 0.3$
 implies that the life-time of a photon of this energy cannot be
much smaller than the age of the Universe (see
Ref.~\cite{Stecker:2001vb} for a  recent discussion of limits
on the Lorentz-violating parameters coming from the observation of
Markarian 501). By making use of the expression
 (\ref{imagE}) for the life-time of  bosons, we see that
this observation excludes theories with the number of extra compact
dimensions smaller than four, just like the proton decay.

The strongest limit may come from the observation of ultra-high energy
cosmic rays. If the recently found correlation~\cite{Tinyakov:2001nr}
of  cosmic rays with energies $E\sim 5\cdot 10^{19}$~eV with BL
Lacertae objects is confirmed by further data, then particles
of this energy should be stable enough to travel as large distances
as $600$~Mpc. By making use of Eqs.~(\ref{imagE}) and (\ref{fwidth})
one finds that the number of extra compact dimensions in our toy model
should be as large as 7, irrespectively of whether primary particles
are
fermions or bosons (assuming that
one can use expressions (\ref{imagE}) and
(\ref{fwidth}) for composite objects like proton). This would mean that
the total number of dimensions is equal to or greater than
12 and
prevent embedding  this brane world into  string/M-theory. Note, however,
that the life-times of  particles strongly depend on the parameter
$k$. Taking this parameter equal to $5M_{Pl}$ will allow to have the
total number of dimensions equal to 11,  a value favored by
M-theory.

To conclude, we constructed a setup with an infinite extra dimension
where bulk space does not exhibit even approximate $SO(1,3)$ isometry
but low-energy effective theory is approximately Lorentz
invariant. This Lorentz invariance is automatic for scalar field,
photon and graviton. Additional fine-tuning of parameters is needed to
assure that the maximum attainable speed of fermions is the same. It
worth noting that with this fine-tuning,
$SO(1,3)$ is still strongly violated the bulk. 


A peculiar property of our setup is that there are no truly localized
modes --- all four-dimensional particles are  narrow
resonances in the spectrum of bulk modes. These resonances have 
finite probabilities to tunnel into  extra dimension.
These probabilities depend on the energies of the particles.

In the (generalized) holographic interpretation where
 bulk modes are described
as a hidden (quasi)conformal sector (see, e.g.,
Refs.~\cite{Gubser:1999vj,Giddings:2000mu,Giddings:2000ay,Arkani-Hamed:2000ds})
this type of decay may
be attributed to the direct coupling 
\[
\phi{\cal O}_{\phi}
\]
between quasilocalized fields $\phi$ and operators ${\cal O}_{\phi}$
from the conformal sector. This coupling may be Lorentz invariant like
in the case of massive particles in the Randall --Sundrum
background~\cite{DRT1} or violating Lorentz invariance like in the
examples of this paper. Tunneling into extra dimension is interpreted
then as a decay into conformal matter.
It worth noting that in this language, emergence
of the Lorentz invariant sector at low energies, in a theory without
Lorentz invariance at high energies, appears rather non-trivial.

Possibly, one
can consider bulk metrics even without $SO(3)$ symmetry under spatial
rotations. For instance, one can take metric with time and one of the
space coordinates  warped and two other coordinates  not warped.
Presumably, one will be able to
 fine-tune parameters so that all maximum attainable
speeds will be equal for different fields in this setup as well. An
interesting property of these metrics is that the tunneling
probability will depend not only on the energy of the particle, but 
also on the
direction of its motion.

The author is indebted to T.~Gherghetta, M.~Giovannini, M.~Libanov,
V.~Rubakov, M.~Shaposhnikov, S.~Sibiryakov, P.~Tinyakov and
S.~Troitsky for helpful discussions of different topics related to
this work. The author
 thanks Institute of Theoretical Physics, University of
Lausanne, where this work was done, for kind hospitality. This work
was supported in part by RFBR grant 99-01-18410, by the Council for
Presidential Grants and State Support of Leading Scientific Schools,
grant 00-15-96626, by CRDF grant (award RP1-2103) and by Swiss Science
Foundation grant 7SUPJ062239.


{\small
\begin{thebibliography}{99}
\bibitem{RS} V.~A.~Rubakov and M.~E.~Shaposhnikov,
Phys.\ Lett.\ B {\bf 125}, 136 (1983).
\bibitem{Akama:1982jy}
K.~Akama,
Lect.\ Notes Phys.\  {\bf 176}, 267 (1982)
[arXiv:hep-th/0001113].
\bibitem{Arkani-Hamed:1998rs}
N.~Arkani-Hamed, S.~Dimopoulos and G.~Dvali,
Phys.\ Lett.\ B {\bf 429}, 263 (1998)
[hep-ph/9803315]; I.~Antoniadis, N.~Arkani-Hamed, S.~Dimopoulos and G.~Dvali,
Phys.\ Lett.\ B {\bf 436}, 257 (1998)
[hep-ph/9804398].
\bibitem{LRS} L.~Randall and R.~Sundrum,
Phys.\ Rev.\ Lett.\ {\bf 83}, 4690 (1999)
[hep-th/9906064].
\bibitem{Volobuev:1986wc}
I.~P.~Volobuev and Y.~A.~Kubyshin,
Theor.\ Math.\ Phys.\ {\bf 68} (1986) 788;
Theor.\ Math.\ Phys.\ {\bf 68} (1986) 885;
JETP Lett.\ {\bf 45} (1987) 581;
I.~Antoniadis,
Phys.\ Lett.\ B {\bf 246}, 377 (1990).
\bibitem{Rubakov:1983bz}
V.~A.~Rubakov and M.~E.~Shaposhnikov,
Phys.\ Lett.\ B {\bf 125}, 139 (1983).
\bibitem{Arkani-Hamed:2000eg}
N.~Arkani-Hamed, S.~Dimopoulos, N.~Kaloper and R.~Sundrum,
Phys.\ Lett.\ B {\bf 480}, 193 (2000)
[hep-th/0001197].
\bibitem{Kachru:2000hf}
S.~Kachru, M.~Schulz and E.~Silverstein,
Phys.\ Rev.\ D {\bf 62}, 045021 (2000)
[hep-th/0001206].
\bibitem{Randall:1999ee}
L.~Randall and R.~Sundrum,
Phys.\ Rev.\ Lett.\ {\bf 83}, 3370 (1999)
[hep-ph/9905221].
\bibitem{GRS} R.~Gregory, V.~A.~Rubakov and S.~M.~Sibiryakov,
Phys.\ Rev.\ Lett.\ {\bf 84}, 5928 (2000)
[hep-th/0002072]; I.~I.~Kogan, S.~Mouslopoulos, A.~Papazoglou, G.~G.~Ross and J.~Santiago,
Nucl.\ Phys.\ B {\bf 584}, 313 (2000)
[hep-ph/9912552];
I.~I.~Kogan, S.~Mouslopoulos and A.~Papazoglou,
Phys.\ Lett.\ B {\bf 503}, 173 (2001)
[hep-th/0011138];
M.~Porrati,
Phys.\ Lett.\ B {\bf 498}, 92 (2001)
[hep-th/0011152];
A.~Karch and L.~Randall,
``Locally localized gravity,''
hep-th/0011156.
\bibitem{DRT2} S.~L.~Dubovsky, V.~A.~Rubakov and P.~G.~Tinyakov,
JHEP{\bf 0008}, 041 (2000)
[hep-ph/0007179].
\bibitem{ST} M.~Shaposhnikov and P.~Tinyakov,
{\it ``Extra dimensions as an alternative to Higgs mechanism?,''}
hep-th/0102161.
\bibitem{Nielsen:1983sz}
H.~B.~Nielsen and I.~Picek,
Nucl.\ Phys.\ B {\bf 211}, 269 (1983)
[Addendum-ibid.\ B {\bf 242}, 542 (1983)].
\bibitem{Chadha:1983qq}
S.~Chadha and H.~B.~Nielsen,
Nucl.\ Phys.\ B {\bf 217}, 125 (1983).
\bibitem{Kostelecky:2000fb}
V.~A.~Kostelecky,
arXiv:hep-ph/0005280.
\bibitem{Bertolami:2000da}
O.~Bertolami and C.~S.~Carvalho,
Phys.\ Rev.\ D {\bf 61}, 103002 (2000)
[arXiv:gr-qc/9912117].
\bibitem{Colladay:1997iz}
D.~Colladay and V.~A.~Kostelecky,
Phys.\ Rev.\ D {\bf 55}, 6760 (1997)
[arXiv:hep-ph/9703464].
\bibitem{Coleman:1999ti}
S.~Coleman and S.~L.~Glashow,
Phys.\ Rev.\ D {\bf 59}, 116008 (1999)
[hep-ph/9812418].
\bibitem{Bear:2000cd}
D.~Bear, R.~E.~Stoner, R.~L.~Walsworth, V.~A.~Kostelecky and C.~D.~Lane,
Phys.\ Rev.\ Lett.\  {\bf 85}, 5038 (2000)
[arXiv:physics/0007049].
\bibitem{Chung:2000ji}
D.~J.~Chung, E.~W.~Kolb and A.~Riotto,
{\it ``Extra dimensions present a new flatness problem,''}
hep-ph/0008126.
\bibitem{Csaki:2000dm}
C.~Csaki, J.~Erlich and C.~Grojean,
{\it 
``Gravitational Lorentz violations and adjustment of the cosmological
constant in asymmetrically warped spacetimes,''} 
hep-th/0012143.
\bibitem{Visser:1985qm}
M.~Visser,
Phys.\ Lett.\ B {\bf 159}, 22 (1985)
[arXiv:hep-th/9910093].
\bibitem{earlier} D.~J.~Chung and K.~Freese,
Phys.\ Rev.\ D {\bf 62}, 063513 (2000)
[hep-ph/9910235];
\bibitem{DRT1} S.~L.~Dubovsky, V.~A.~Rubakov and P.~G.~Tinyakov,
Phys.\ Rev.\ D {\bf 62}, 105011 (2000)
[hep-th/0006046].
\bibitem{Gherghetta:2000jf}
T.~Gherghetta, E.~Roessl and M.~Shaposhnikov,
Phys.\ Lett.\ B {\bf 491}, 353 (2000)
[hep-th/0006251].
\bibitem{Ponton:2001gi}
E.~Ponton and E.~Poppitz,
JHEP{\bf 0102}, 042 (2001)
[hep-th/0012033].
\bibitem{Dvali:1997xe}
G.~Dvali and M.~Shifman,
Phys.\ Lett.\ B {\bf 396}, 64 (1997)
[hep-th/9612128].
\bibitem{Bajc:2000mh}
B.~Bajc and G.~Gabadadze,
Phys.\ Lett.\ B {\bf 474}, 282 (2000)
[hep-th/9912232].
\bibitem{Randjbar-Daemi:2000cr}
S.~Randjbar-Daemi and M.~Shaposhnikov,
Phys.\ Lett.\ B {\bf 492}, 361 (2000)
[hep-th/0008079].
\bibitem{Jackiw:1976fn}
R.~Jackiw and C.~Rebbi,
Phys.\ Rev.\ D {\bf 13}, 3398 (1976).
\bibitem{Brandhuber:1999hb}
A.~Brandhuber and K.~Sfetsos,
JHEP{\bf 9910}, 013 (1999)
[hep-th/9908116].
\bibitem{Neronov:2001br}
A.~Neronov,
{\it ``Localization of Kaluza-Klein gauge fields on a brane,''}
hep-th/0102210.
\bibitem{Lykken:2000nb}
J.~Lykken and L.~Randall,
JHEP{\bf 0006}, 014 (2000)
[hep-th/9908076].
\bibitem{Giddings:2000mu}
S.~B.~Giddings, E.~Katz and L.~Randall,
JHEP{\bf 0003}, 023 (2000)
[hep-th/0002091].
\bibitem{Dvali:2000rv}
G.~R.~Dvali, G.~Gabadadze and M.~Porrati,
Phys.\ Lett.\ B {\bf 484}, 112 (2000)
[arXiv:hep-th/0002190].
\bibitem{Deffayet:2001uk}
C.~Deffayet, G.~R.~Dvali, G.~Gabadadze and A.~I.~Vainshtein,
arXiv:hep-th/0106001.
\bibitem{Groom:2000in}
D.~E.~Groom {\it et al.}  [Particle Data Group Collaboration],
Eur.\ Phys.\ J.\ C {\bf 15}, 1 (2000).
\bibitem{mark} F.A.~Aharonian {\it et. al.}, Astron. and Astrophys.,
{\bf 349}, 11 (1999), arXiV:astro-ph/9903386.
\bibitem{Stecker:2001vb} F.~W.~Stecker and S.~L.~Glashow, {\it ``New
tests of Lorentz invariance following from observations of the highest
energy cosmic gamma rays,''} astro-ph/0102226.
\bibitem{Tinyakov:2001nr}
P.~G.~Tinyakov and I.~I.~Tkachev,
{\it ``BL Lacertae are sources of the observed ultra-high energy
cosmic rays,''} 
astro-ph/0102476.
\bibitem{Gubser:1999vj}
S.~S.~Gubser,
{\it ``AdS/CFT and gravity,''}
hep-th/9912001.
\bibitem{Giddings:2000ay}
S.~B.~Giddings and E.~Katz,
{\it ``Effective theories and black hole production in warped
compactifications,''} 
hep-th/0009176.
\bibitem{Arkani-Hamed:2000ds}
N.~Arkani-Hamed, M.~Porrati and L.~Randall,
{\it ``Holography and phenomenology,''}
hep-th/0012148.
\end{thebibliography}}
\end{document}